\documentclass[prl,twocolumn,showpacs]{revtex4}
\usepackage{graphics}
\begin{document}

\title{A new mechanism for a naturally small Dirac neutrino mass}
\author{P.Q. Hung}
\email[]{pqh@virginia.edu}
\affiliation{Dept. of Physics, University of Virginia, \\
382 McCormick Road, P. O. Box 400714, Charlottesville, Virginia 22904-4714}
\date{\today}
\begin{abstract}
%A new mechanism for a small neutrino mass is presented in this
%paper. It relies on the idea that the size of the overlap
%between the wave functions of left-handed and right-handed
%fermion zero modes (quarks and leptons) along
%an extra compact dimension practically ``determines''
%the value of the effective Yukawa coupling in four dimensions.
%It is shown here how neutrino and charged lepton 
%right-handed zero modes can, 
A mechanism is proposed in which a right-handed neutrino zero mode
and a right-handed charged lepton zero mode can be localized 
at the same place along an extra compact dimension while
having markedly different spreads in their wave functions: a
relatively narrow one for the neutrino and a rather broad one for
the charged lepton. In their overlaps with the wave function
for the left-handed zero modes, this mechanism could produce
a natural large hierarchy in the effective Yukawa couplings 
in four dimensions, and hence a large disparity in masses.
\end{abstract}
\maketitle
The possibility that a neutrino has a small non-vanishing mass
(of order of a few electronvolts or less) is
by now universally accepted and backed by various neutrino
oscillation experiments \cite{osc}. However what is not established at
this moment is the nature of the mass itself. Is it of the
Dirac type or is it of the Majorana type? There are ongoing and 
planned experiments designed to address this very issue
\cite{onubeta, katrin}. Despite the
general belief that the masses of the neutrinos are of the
Majorana type, it is prudent to keep in mind that they
could be of the pure Dirac type and to explore scenarios
pertaining to this latter possibility. The final words
must evidently come from experiments. 

%(SHORTEN THIS PART)
The common prejudice in favor of a Majorana mass is a very natural
one. The see-saw mechanism \cite{seesaw}
with a ratio of a small scale (electroweak)
/a large scale (Grand Unified scale, etc..) provides a
plausible explanation for the smallness of the Majorana
neutrino mass. 
%(The connection with the quark sector is
%however less evident.)
%Accordingly, the smallness of the neutrino masses is derived
%from the famous see-saw mechanism and this can be 
%realized in a number of extensions of the Standard Model (SM)
%such as the left-right symmetric model, $SO(10)$, etc..
%(The inclusion of three families and their mixings is
%another matter.) In
%the see-saw mechanism, the ratio of a neutrino mass over that
%of its charged lepton counterpart is given typically by the
%ratio of the electroweak scale over some GUT (Grand Unified Theory)
%scale (keeping in mind that there are variants of this simple
%scenario), and it could be very small.
If one simply enlarges the minimal SM by 
endowing it with right-handed neutrinos, 
the neutrino masses would
naturally be of the Dirac type and $m_{\nu}/
m_{l} = g_{Y\nu}/g_{Yl} \alt 10^{-9}$.
It is then considered to be highly unnatural
(fine tuning)  to put in
by hand $g_{Y\nu} \alt O(10^{-11})$ if $g_{Yl} \sim 10^{-2}$
(roughly the value of the Yukawa coupling of the heaviest
charged lepton).
It would
be more natural if $g_{Y\nu} \alt O(10^{-11})$ were to
arise dynamically. This was the motivation of Ref. 
\cite{hung1} where Dirac neutrino masses arise at the
one-loop level and can be ``naturally'' small.
%In fact, t
%Through
%a simple approximation, these masses are also
%look very similar to
%those obtained by the see-saw mechanism in that they are
%proportional to a ratio of a small scale over a large scale.
%One can, of course, debate on the merits of various scenarios
%and the physical meanings of the two (or more) scales. However, a 
%A generic feature seems to emerge from these works: the
%smallness of neutrino masses come from the large disparity
%between two different mass scales.

In this paper, another approach concerning a naturally
small Dirac mass for the neutrinos is presented.
%The proposed mechanism is also applied to
%the quark sector with intriguing implications.
A very interesting connection to the quark 
sector will be shown.
It makes use of the interesting notion that the effective
Yukawa coupling (in four dimensions) 
controlling in part the value of the fermion
mass is proportional to the size of the wave function overlap 
between left-handed and right-handed fermions along the compact
fifth (spatial) dimension. 

Let us start with the case where there is one extra
spatial dimension- denoted by $y$- compactified 
on an orbifold $S_{1}/Z_{2}$ and having length $L$. This
has been shown to contain chiral zero modes in four
dimensions \cite{georgi:2000wb}, a desirable feature in the construction
of the SM. These chiral zero modes
(for either left or right-handed mode) can be written as
$\psi_{L,R}^{0}(x, y) = \psi_{L,R}(x)
\xi_{L,R}(y)$, where
$x \equiv x^{\mu}$. For free fields, $\xi_{L,R}(y)$ = 
constant and the wave function is uniformly spread
over the extra dimension. When a background scalar
field with a kink solution is introduced and is 
made to interact with the fermion, it was found that 
the zero mode is localized at the center of the domain
wall (the zero of the kink) and rapidly vanishes
outside it. 
%The location of the domain wall
%itself is a parameter which presumably is determined at
%a deeper level. 
As it has been discussed in Ref. \cite{arkani:1999dc}, one
can arrange domain walls at various locations 
along $y$ so as to place fermions at different points.
%Various phenomenological implications were made
%along this line.

What has not been discussed so far is the possibility
that each fermion interacts with more than one background scalar
field and the ensuing implications on the shape and location of
the wave function along $y$. This turns
out to be very relevant to the
issue of naturally small Dirac neutrino masses, as
well as to the splitting between the up and down sectors
of the quarks.

Let us first start with the simple case of one fermion, $\psi$,
interacting with two background scalar fields, $\phi_1$ and
$\phi_2$, as follows
\begin{equation}
\label{yuk1}
{\cal L}_{Y}= f_{1}\bar{\psi}\psi\,\phi_{1} +
f_{2}\bar{\psi}\psi\,\phi_{2} \,.
\end{equation}
For simplicity, let us take the non-vanishing chiral
zero mode to be a right-handed one. This can be
accomplished by choosing the appropriate $Z_2$ parity.
(The other chirality can be treated in the same way.)
As in Ref. \cite{georgi:2000wb, arkani:1999dc}, kink solutions 
will be assumed for $\phi_1$ and
$\phi_2$, which will be denoted by $h_{1}(y)$ and
$h_{2}(y)$ and are given by 
\label{kink}
\begin{equation}
h_{1,2}(y) = v_{1,2} \tanh\{\mu_{1,2}\,(y-y_{1,2})\} \, ,
\end{equation}
where $\mu_{1,2} = (\lambda_{1,2}/2)^{1/2} v_{1,2}$ 
($\lambda_{1,2}$ being the coefficients of the quartic
couplings), and where we have allowed for the locations
of the zeros of the kinks to be at $y_{1,2}$. 
The equation for $\xi_{R}$ is given by
\begin{equation}
\label{eqxi}
\partial_{y}\xi_{R}(y)+(f_{1}h_{1}(y) +
f_{2}h_{2}(y))\xi_{R}(y) = 0\,,
\end{equation}
The normalized zero mode wave function for the fermion is
now given by
\begin{equation}
\label{xi}
\xi_{R}(y) = k \exp(-\int_{0}^{y} dy^{\prime} (f_{1} h_{1}(y^{\prime})+
f_{2} h_{2}(y^{\prime}))) \, ,
\end{equation}
%where
%\begin{equation}
%\label{s(y)}
%s(y) = \int_{0}^{y} dy^{\prime} (f_{1} h_{1}(y^{\prime})+
%f_{2} h_{2}(y^{\prime}))\, ,
%\end{equation}
where $k$ is a normalization factor so that
$\int_{0}^{L}\xi_{R}^2 =1$. To illustrate
the above discussion, let us, for simplicity, set
$y_{1} = y_{2} =0$ and call $f_{1,2}/(\lambda_{1,2}/2)^{1/2}=
C_{1,2}$. It can easily be seen that $\xi_{R}(y)$ now takes
the simple form
\begin{equation}
\label{xi2}
\xi_{R}(y) = k e^{-C_{1} \ln(\cosh(\mu_{1}y))-
C_{2} \ln(\cosh(\mu_{2}y))} \, .
\end{equation}

To illustrate (\ref{xi2}) graphically, we shall simply
put $C_{1}=\pm 1$, $C_{2} = \pm 1$, with the
appropriate normalization factor $k$ 
and, in an arbitrary
unit of masses, $\mu_{1}=1$, $\mu_{2} = 0.9$. 
(A more general case can be straightforwardly
carried out.) Since
$1/\mu$ corresponds to the thickness of the domain
wall, these illustrative numerical values correspond
to two domain walls of comparable thicknesses.
(However, this is {\em not} ``fine tuning''.)
In Fig. 1, we show two cases (the sum and
the difference respectively): (a) $C_{1} =1$, 
$C_{2}=1$; (b) $C_{1} =1$, $C_{2}=-1$. These two 
cases correspond to two different wave functions
which are denoted by WFa and WFb respectively. 
Also shown are the two kinks located at the same point
but endowed with slightly different profiles. 
%orthe sake of discussion and illustration, 
We also show another
fermion wave function (e.g. a left-handed one),
denoted by WFL which is 
%which does not interact with the above two background
%scalar fields and which is chosen to be 
localized separately at
a point far enough from the two domain walls.
The domain wall with which it interacts is chosen
to be of comparable size to the other two.
 
From Fig. 1, one can see that WFa
is markedly different from WFb.
It is much narrower while the latter is rather
broad. (The difference in the maxima of WFa and WFb
reflects the two different normalization factors
which are computed numerically here.)
For illustrative purposes, the (left-handed)
wave function, WFL, on the left of Fig. 1 is chosen to
have a tiny overlap with WFa. 
However, its overlap with WFb is large, due to the spread of WFb. 
%For the sake of comparison, Cases (a) and (b) are also
%shown in Fig. 1 for $\mu_{1}=1$ and $\mu_{2} = 0.1$ (the
%second domain wall is ten times thicker than the first
%one). One can then see that WFb $\sim$ WFa and this
%clearly illustrates the phenomenon that the spreading
%of WFb is related to the situation when the two
%domain walls are comparable in thickness, i.e. $\mu_{1}
%\sim \mu_{2}$.
%Or, put it another way, the location of WFL is chosen
%so that it has a ``large'' overlap with WFb. As a
%consequence, its overlap with WFa is small.
In the parlance of effective couplings in four dimensions,
WFL-WFa and WFL-WFb overlaps would give rise to two
markedly different effective couplings: a tiny one and
a much larger one respectively. If these were the Yukawa
couplings responsible for the masses, one would have
obtained two widely separated masses. One could
envision the possibility that the (right-handed)
fermion involved in the WFL-WFa overlap is the
right-handed neutrino and the one involved in the
WFL-WFb overlap is the (right-handed) charged lepton.
But how does one realize this scenario in a realistic 
model?

Although we have alluded to a scenario where the
aforementioned fermions are leptons, one
can also imagine a similar one for the quarks,
with a notable difference: The split of the
overall mass scales between the up and down sector
is about a factor of 60 or so. In this case, one
might imagine moving WFL closer to the center
of WFa and WFb, and reverse the coupling of the
right-handed quarks (coupling to the sum for the
down quark (WFa) and to the difference for the up quark
(WFb),
all to the same background scalar fields).
Fig. 2 illustrates this scenario.  
By choosing the appropriate location for WFL, one can
reproduce the above desired ratio. It is
intriguing that, with the {\em same} two background
scalars, the ratio of the overall mass scales
between the up and down sectors for {\em both} leptons
and quarks can be obtained by choosing the
appropriate {\em relative} (to the right-handed fermions)
distance for WFL(lepton) and WFL(quark). (Only
these relative distances are relevant here.)
%As in Ref. (\cite{arkani:1999dc}), one can shift the locations
%of those same kinks by adding ``mass'' terms
%(in the sense of Ref. (\cite{arkani:1999dc})
%to the quarks so that those positions are
%determined by the zeros of $f\,h(y)-m$. The 
%reason for so doing is to prevent any
%significant overlap between the quark and lepton
%wave functions. As a cautionary remark, the
%coordinates shown in Fig.2 are for illustration
%and are not the same as those of Fig.1.
For simplicity, the ``origin'' in Fig. 1 and 2 really means
the relative location of the right-handed fermions
and not their actual positions along $y$.

To summarize, given two
background scalar fields and two different fermions,
the coupling of one fermion to the sum of the scalars
(Case (a)) gives rise to a ``narrow'' localized wave 
function while the one which couples to the difference 
(Case (b)) gives rise to a ``broad'' localized 
wave function. 
%A more general case keeping $C_1$,
%$C_2$, 
Similar conclusions can be reached for other values of
$\mu_1$ and $\mu_2$ as long as they are not too different 
from one another.
%arbitrary is straightforward
%to carry out. 

%(THIS PART IS TO BE WRITTEN VERY CAREFULLY.)
%Before describing the actual construction, a few remarks are
%in order. The actual locations along $y$ of the left-handed
%and right-handed zero modes depend on a more complete
%theory. One can imagine the following scenario. The
%right-handed neutrino and charged lepton zero modes
%are localized at some point along $y$, as shown in Fig. 1.
%The interaction with the two background scalars splits
%the two wave functions as in Cases (a) and (b). Then WFL
%is positioned so that its overlap with WFb produces the
%``correct'' charged lepton mass scale. The domain wall 
%thicknesses are then adjusted so as the ``correct''
%neutrino mass scale appears. Or- and this might be heretic 
%thought- one can imagine that, somehow, there is a
%mechanism which puts the right-handed lepton zero modes
%far apart from its left-handed counterpart so that
%both neutrinos and charged leptons would have very small
%masses, if not for the presence of the triplet
%background scalar field which broadens the right-
%handed charged lepton zero mode wave function and
%greatly raises its mass.

Notice that a fermion in five dimensions
is a Dirac fermion which can be written as $\psi = 
(\psi_L+ \psi_R)$, where $\psi_{L,R} = P_{L,R}\psi$, with
$P_{L,R} = (1\mp \gamma_5)/2$ being the usual
four-dimensional chiral projection operator. The $Z_2$ symmetry
in conjunction with the boundary conditions will select one
of the two chiralities to have a non-vanishing zero
mode, depending on the chosen $Z_2$ parity. 
%The other one will
%be its heavy (of order $1/L$) mirror counterpart. In
%this sense, all SM chiral fermions will have their
%opposite chirality mirror heavy counterparts.
In addition to the SM, let us introduce an extra symmetry
$SU(2)_{R}$ which could either be global or 
gauged as in the L-R model \cite{LR}.
The notations used for the lepton doublets in four dimensions
are as follows: $l_{L} = (\nu_{L}, e_{L})$ 
($SU(2)_L$ doublet) and $l_{R} = (\nu_{R}, e_{R})$
($SU(2)_R$ doublet), where $\nu$ and $e$ are
generic names for the neutral and charged leptons.
Let us now embed these leptons in five-dimensional spinors,
namely $L^{\{L\}}(x,y) = (l^{\{L\}}_{L}+
l^{\{L\}}_{R})$ and $L^{\{R\}}(x,y) = 
(l^{\{R\}}_{L}+l^{\{R\}}_{R})$. Here, $l_L(x)$ is 
the zero mode of $l^{\{L\}}_{L}(x,y)$, and 
$l_R(x)$ is the zero mode of $l^{\{R\}}_{R}(x,y)$.
In the absence of interactions, these zero modes depend
only on $x$ and are not localized along $y$. 

Let us now concentrate on $L^{\{R\}}$. Let us introduce
two background scalar fields:
$\Phi_{T}= \vec{\phi}_{T}.\frac{\vec{\tau}}{2}$ and $\phi_{S}$,
which are a triplet and a singlet under $SU(2)_R$ respectively,
and hence the subscripts $T$ and $S$. 
%Let us furthermore
%assume that there is an {\em additive} quantum number
%for the leptons and for the background scalar fields as follows:
%+1 for the right-handed leptons,
%-1 for the left-handed leptons and -2 for those scalars, so that
%they do not couple to the quarks. (The assignment for the
%left-handed leptons will be made clear at the end.)
%(CAREFUL HERE!MIGHT NOT NEED THIS QUANTUM NUMBER)
Furthermore,
as usual, these background scalars are assumed to transform
under $Z_2$ as $\phi_{S}(x,y) \rightarrow -\phi_{S}(x,L-y)$,
$\Phi_{T}(x,y) \rightarrow -\Phi_{T}(x,L-y)$ as in Ref. \cite
{georgi:2000wb}.
The Yukawa coupling (in five dimensions) can now be written
as
\begin{equation}
\label{yuk2}
{\cal L}_{Y2}= f_{T}^{(l)}\bar{L}^{\{R\}}\Phi_{T}\,L^{\{R\}} +
f_{S}^{(l)}\bar{L}^{\{R\}}\phi_{S}\,L^{\{R\}} \,.
\end{equation}
With $\Phi_{T}$ being an $SU(2)_R$ triplet, its minimum
energy configuration can be written as
$\langle \Phi_{T} \rangle = \langle\phi_{T}^{3}\rangle
\tau_{3}/2$, and explicitely as
\begin{equation}
\label{PhiT}
\langle \Phi_{T} \rangle = 
\left(\begin{array}{cc}
h_{T}(y)&0 \\
0&-h_{T}(y)
\end{array}
\right)\ \, .
\end{equation}
For $\phi_S$, one has $\langle \phi_{S} \rangle = h_{S}(y)$.
Let us write the zero mode for $l^{\{R\}}_{R}(x,y)$ as
$l^{0,\{R\}}_{R}(x,y) = l_{R}(x)\xi_{R}(y)$. The
equation governing the behaviour of $\xi_R$ is now
\begin{equation}
\label{eqxi2}
\partial_{y}\xi_{R}(y)+(f_{T}^{(l)}\langle \Phi_{T} \rangle +
f_{S}^{(l)}\langle \phi_{S} \rangle)\xi_{R}(y) = 0\,.
\end{equation}
With $\xi_{R} = (\xi^{\nu}_{R}, \xi^{e}_{R})$, (\ref{eqxi2})
splits into
\begin{mathletters}
\begin{equation}
\label{nu}
\partial_{y}\xi^{\nu}_{R}(y)+(f_{S}^{(l)}h_{S}(y) +
f_{T}^{(l)}h_{T}(y))\xi^{\nu}_{R}(y) = 0\,,
\end{equation}
\begin{equation}
\label{e}
\partial_{y}\xi^{e}_{R}(y)+(f_{S}^{(l)}h_{S}(y)
-f_{T}^{(l)}h_{T}(y))\xi^{e}_{R}(y) = 0\,,
\end{equation}
\end{mathletters}
The solutions are
\begin{equation}
\label{xinu}
\xi_{R}^{\nu,e}(y)= k_{\nu,e}\exp(-\int_{0}^{y}dy^{\prime} 
(f_{S}^{(l)} h_{S}(y^{\prime})\pm f_{T}^{(l)} h_{T}(y^{\prime}))\,,
\end{equation}
where $k_{\nu}$ and $k_{e}$ are normalization factors
and where $f_{S,T}$ are assumed to be positive.
Eq. (\ref{xinu}) has the same form as Eq. (\ref{xi}).
The analysis is identical to the generic case
discussed above when one makes identifications of various
coefficients, namely $1,2 \rightarrow S,T$ and $k \rightarrow
k_{\nu,e}$. As for the quark sector, one would like the
right-handed Up wave function to be ``broad'' and the
right-handed Down wave function to be ``narrow''. This
can be accomplished by writing the following term $-f_{T}^{(q)}
\bar{Q}^{\{R\}}\,\Phi_{T}\,Q^{\{R\}} +
f_{S}^{(q)} \bar{Q}^{\{R\}}\,\phi_{S}\,Q^{\{R\}}$,
with $f_{T,S}^{(q)}$ assumed to be positive. With
this, $\xi_{U}(y)$ will behave like $\xi_{e}(y)$, and
$\xi_{D}(y)$ like $\xi_{\nu}(y)$. 
Eqs. (\ref{xinu}) is seen to be an explicit
realization of the phenomenon presented in this paper.
%namely a ``narrow'' localized wavefunction ($\xi_{\nu}$) for
%the interaction with the {\em sum} of the scalars, and
%a ``broad'' localized wave function for the interaction
%with the {\em difference} ($\xi_{e}$). 
%Since the gauge
%symmetry is assumed to be similar to the LR model,

In the following discussion, we will ignore
fermion mixings and concentrate on the overall
mass scales, which usually mean the heaviest
fermion for each sector.
A Yukawa interaction responsible for the masses would
be of the form $g_{Y,l}\bar{L}^{\{L\}}H\,L^{\{R\}}+h.c.$,
where $H$ is a Higgs field, and similarly
for the quarks. If $SU(2)_R$ were a global
symmetry then $H$ would be a $2\times2$ matrix of the
form $H=(\phi, \tilde{\phi})$ with $\tilde{\phi}=i
\sigma_{2}\phi^{\ast}$. Assuming an even $Z_{2}$-parity 
for $H$ so that it has a non-trivial zero mode $H^{(0)}$,
it follows that $\langle H^{(0)} \rangle =diag(v, -v)/
\sqrt{2}$ with $v=246$ GeV. If $SU(2)_R$ were a gauge
symmetry as in the LR model then
$\langle H^{(0)} \rangle =diag(v_1, -v_2)$ with
$v_{1}^2 + v_{2}^2 = 246$ GeV.
For simplicity, we will first concentrate on the
global $SU(2)_R$ case and see what we can 
learn from there. In this case, one only has to deal
with one fundamental scale $v$.
If we denote
by $\xi_{L}$ the corresponding wave function for the
left-handed zero mode, the effective Yukawa couplings
in four dimensions are
\begin{equation}
\label{geff}
g_{eff,\nu} = g_{Y,l} \int_{0}^{L} dy \xi^{l}_{L} \xi_{R}^{\nu}\,;\,
g_{eff,e} = g_{Y,l} \int_{0}^{L} dy \xi^{l}_{L} \xi_{R}^{e}\,,
\end{equation}
and similarly for the quark couplings, $g_{eff,U}$ and
$g_{eff,D}$. The overall mass scales will then be:
$g_{eff,\nu}\,v/\sqrt{2}$ and $g_{eff,e}\,v/\sqrt{2}$
for the neutrino and charged lepton sectors respectively,
and $g_{eff,U}\,v/\sqrt{2}$ and
$g_{eff,D}\,v/\sqrt{2}$ for the Up and Down quark sectors
respectively. In this scenario, the ratio of the
two mass scales for the leptons and for the quarks
are simply ratios of the overlaps. The fundamental
Yukawa couplings $g_{Y,l,q}$ will come in when one 
wants to set the absolute scales of the masses.

%The above discussion is an explicit model for the phenomena
%discussed at the beginning of this paper. An
%intriguing behaviour occurs and this is illustrated
%in Figs. (1,2). 
Although our original motivation
was to find a natural reason for a small Dirac
neutrino mass, one can turn the argument around
and apply the proposed mechanism to the quark sector
first followed by a ``prediction'' for the lepton
sector. Basically,
the main philosophy of \cite{arkani:1999dc} 
is that the effective Yukawa couplings
$g_{eff}$ can be small even if the {\em fundamental}
Yukawa couplings $g_{Y}$ were of O(1). 
%We will
%exploit this philosophy in order to make a connection
%between the quark and lepton sectors below.
%(One might
%argue that the problem is reshuffled into the size
%of the overlap which, however, could be viewed as
%a dynamical quantity.) 
%In the minimal SM,
%the heaviest charged lepton ($\tau$) has a Yukawa
%coupling $g_{Y} \sim 10^{-2}$. For that matter,
%even the bottom quark Yukawa coupling is small.
Let us assume here- for
the sake of discussion- that all fundamental
Yukawa couplings $g_{Y,l,q} \gtrsim 0.1$. (There
is no deep reason for this lower bound other than
a generous definition of what O(1) means.) Let
us illustrate our scenario with the parameters chosen
for Figs. (1,2). (A general discussion is beyond
the scope of this paper.) From Fig. 2, the ``broad''
wave function represents $U_R$ and the narrow one for $D_R$.
Since $g_{Y,q}v/\sqrt{2}$ is common, $m_t/m_b$ is just
the ratio of overlaps and is about 60. The actual
value for $m_t(m_t) \sim 166$ GeV yields $g_{Y,q}
\sim 2.5$. (Remember that, in four dimensions, 
it is $g_{eff,U}$ which is used in perturbation theory
and, in this particular case, is about unity.)
It is the same two kinks which interact with the leptons,
albeit at a different location. We then move to Fig. 1,
where the origin now refers to the location of the
right-handed doublet. We can then move the left-handed
doublet (WFL) to the left {\em until} the fundamental coupling
$g_{Y,l} \sim 0.1$ (from $m_{\tau}$). 
There, $m_{\nu}/m_{\tau} \sim 10^{-7}$.
The neutrino mass is {\em very small}, albeit one with
the presently-believed wrong value ($\sim 160$ eV).
Moving futher to the left until $m_{\nu}/m_{\tau} \sim 2
\times 10^{-9}$, one gets $g_{Y,l} \sim 0.16$ and
$m_{\nu} \sim 4.4$ eV. It is amusing to push further to
the left until $g_{Y,l} \sim 1$ giving $m_{\nu}/m_{\tau} \sim 2
\times 10^{-16}$ and $m_{\nu} \sim 4\times 10^{-7}$ eV.
This last example shows that the neutrino mass is automatically
very small when $g_{Y,l} \sim g_{Y,q}$ and when one fixes 
the overlap $\int_{0}^{L} dy \xi^{l}_{L} \xi_{R}^{e}$ so that 
the charged lepton acquires a mass of the right magnitude.

From the above examples, one cannot fail but notice a deep
connection between quark and lepton masses. The ratios of
masses for each sector are basically ratios of
effective Yukawa couplings, namely $R_{L} \equiv
g_{eff,\nu}/g_{eff,l}$ and $R_{Q} \equiv
g_{eff,D}/g_{eff,U}$. The only difference between these
ratios is the distances between WFL and WFa,b for the
quark ($d_{q}$) and lepton ($d_{l}$) sectors, and they might come from
a deeper theory unifying quarks and leptons. What this
might mean is that, if such a unification exists, one might
be able to relate these two distances so that using
e.g. $d_{q}$ to calculate $R_{Q}$, one can deduce $R_{l}$
since $d_{l}$ and $d_{q}$ will then be related.
%One advantage of dealing with these
%ratios is the fact that the fundamental Yukawa couplings
%cancel out. Furthermore, in the ratio $R_{L}/R_{Q}$, the
%normalization factors cancel out as well. 
%If we denote
%by $d_{l}$ and $d_{q}$ the distances along $y$ between
%WFL and WFa,b for the lepton and quark sectors respectively
%(Fig. 1 and 2),
%a back-of-the-envelope argument gives
%$R_{L}/R_{Q} \sim \exp(-\mu_{T}^{2}(d_{l}^2 -d_{q}^2))$.
What does the above numerical example tell us? The
mechanism which splits the right-handed doublet into
a ``broad'' and a ``narrow'' wave functions fixes the
ratio of the Up and Down masses for the quarks as well as
for the leptons and naturally gives rise to a tiny Dirac
neutrino mass if one requires that the fundamental Yukawa
couplings (in the higher dimensions) are of O(1).

This paper is concentrated mainly on a mechanism for the disparity
of the overall mass scales, and has not discussed the issue of
mixings. This will be carried out in a separate publication for
the lepton sector. The mass matrices for the quark sector were
investigated in \cite{hung2}.

\begin{acknowledgments}
This work is supported in parts by the US Department
of Energy under grant No. DE-A505-89ER40518. I would like to thank
Sally Dawson for discussions and the High Energy Theory Group
at Brookhaven National Laboratory for hospitality where part
of this work ws carried out. I would also like to thank G. Pancheri
for the hospitality at LNF(Frascati).
\end{acknowledgments}

% now the references. delete or change fake bibitem. delete next three
%   lines and directly read in your .bbl file if you use bibtex.

\begin{figure}
\includegraphics{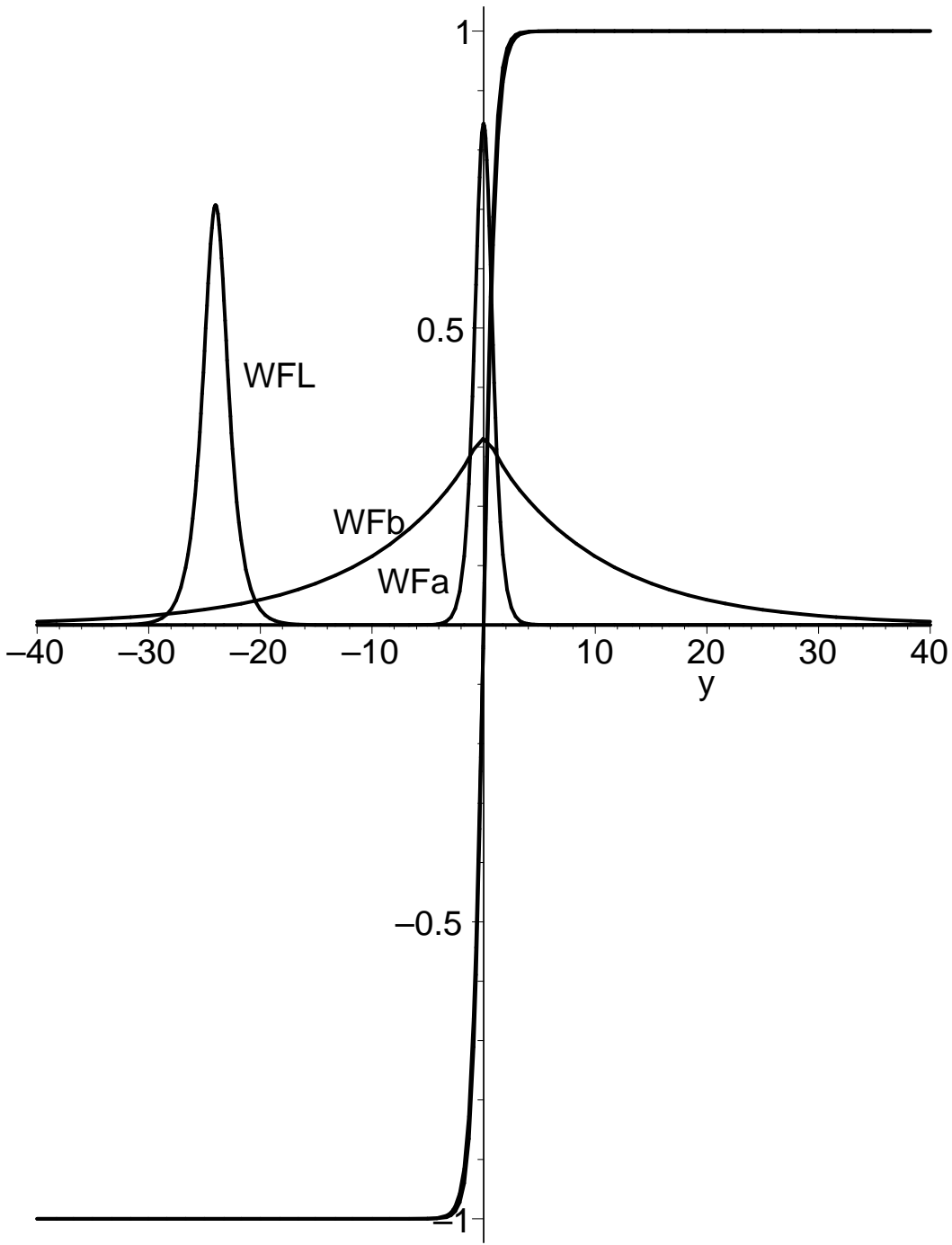}
\caption{\label{fig1}The wave function (WF) overlap for the lepton case.
WFa: right-handed neutrino WF. WFb: right-handed charged lepton WF.
WFL: WF of left-handed  lepton doublet. The two kinks are practically on top
of one another.}
\end{figure}

\begin{figure}
\includegraphics{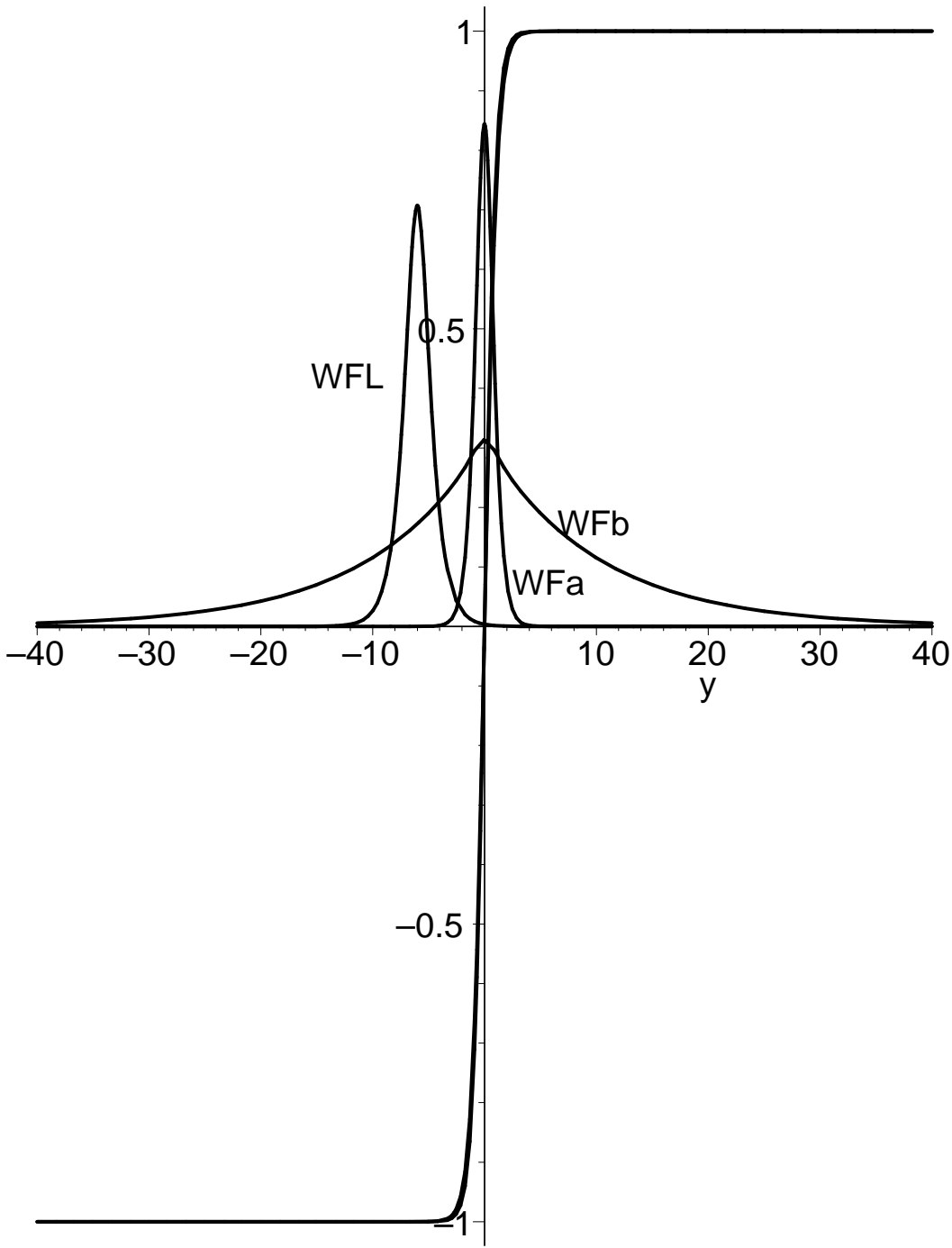}
\caption{\label{fig2}The wave function (WF) overlap for the quark case.
WFa: right-handed down quark WF. WFb: right-handed up quark WF.
WFL: WF of left-handed quark doublet. The two kinks are practically on top of one another.}
\end{figure}

% \begin{table}
% \caption{}
% \label{}
% \end{tabular}
% \end{table}

\end{document}